\documentstyle[aps,epsf]{revtex}
\begin{document}
%
%
%
\begin{center}
  {\Large\bf Parametric Amplification of Density Perturbation in }\\
  {\Large\bf the Oscillating inflation}
\end{center}
\centerline{  
  \large A.Taruya
  \footnote{e-mail:~ataruya@vishnu.phys.h.kyoto-u.ac.jp}
  }
\bigskip
\begin{center}
{\em Department of Fundamental Sciences, FIHS, Kyoto University, 
Kyoto, 606-8501, Japan }
\end{center}
\begin{center}
  {\large Abstract}
\end{center}
We study the adiabatic density perturbation in 
the {\it oscillating inflation}, proposed by Damour and Mukhanov. 
The recent study of the cosmological perturbation during 
reheating shows that the adiabatic fluctuation 
behaves like as the perfect fluid and no significant amplification 
occurs on super-horizon scales.  
In the oscillating inflation, however, the accelerated expansion 
takes place during the oscillating stage and 
there might be a possibility that the parametric amplification on 
small scales affects the adiabatic long-wavelength perturbation.  
We analytically show that 
the density perturbation neglecting the metric perturbation can be 
amplified by the parametric resonance and the instability 
band becomes very broad during the oscillating inflation. 
We examined this issue by solving the evolution equation 
for perturbation numerically. We found that the parametric 
resonance is strongly suppressed 
for the long wave modes comparable to the Hubble horizon. 
The result indicates that the metric perturbation plays 
a crucial role for the evolution of scalar field perturbation. 
Therefore, in the single field case, 
there would be no significant imprint of the oscillating inflation 
on the primordial spectrum of the adiabatic perturbation. However, 
it could be expected that the oscillating inflation in the 
multi-field system gives the enormous amplification on large scales, 
which may lead to the production of the primordial black holes. 
%
%
\section{Introduction}
%
%
%
  Dynamics of the coherently oscillating scalar field 
  play an important role in the early stage of 
  the universe and has been studied by many authors. 
Specifically, the significance of parametric resonance 
  is found in the recent investigation of the reheating 
  process\cite{KLS94}. Due to the coherent oscillation of 
  the background inflaton field, the light boson field or 
  the fluctuation of the inflaton itself is amplified 
  through the non-linear interaction or the self-interaction
  \cite{KLS97,GKLS97}. 
  The efficiency of the parametric resonance could have 
  the important implication for GUT scale baryogenesis\cite{KLR96}.

From the viewpoint of the large scale structure in the early universe, 
several authors investigated  
the influence of parametric resonance on the 
primordial density perturbation during the coherently oscillating stage 
after inflation. In the framework of cosmological perturbation, 
the single field model is studied in Ref.\cite{KH96,NT97}.   
The analysis is also extended to the two-field model by Taruya and 
Nambu \cite{TN98}. These analyses are mainly focused on the 
long wavelength perturbation, which significantly differs from the 
one neglecting the metric perturbation. 
Nambu and Taruya \cite{NT97} investigate the Mukhanov's gauge-invariant 
variable and found that the evolution equation for perturbation 
can be reduced to the Mathieu type equation. Although the Mathieu 
equation itself has the exponential instability, the instability of 
the perturbation is very different from this and the final result is 
fully consistent with Kodama and Hamazaki \cite{KH96}. 
The important conclusion of these analyses is that the density perturbation 
on super-horizon scales behaves like as the perfect fluid unless the 
iso-curvature perturbation 
becomes dominant. Therefore, as far as the adiabatic perturbation 
is concerned, the primordial power spectrum on large scales does not 
suffer from the significant amplification by the parametric resonance.  
However, there remains a possibility that the parametric resonance 
can appear well inside the Hubble horizon. For example, consider 
the massless self-interacting inflaton. Neglecting the metric 
perturbation, it is known that the fluctuation of the inflaton field 
has the instability mode whose wavelength is much smaller than the 
Hubble horizon \cite{KLS94,GKLS97}.  

The aim of the present paper is to clarify the 
perturbaton on sub-horizon scales taking into account 
the gravitational perturbation and 
understand the cosmological implication of the parametric amplification 
to the large scale structure. 
  As usual, the universe expands decelerately during the coherently 
  oscillating stage and the parametric resonance inside the 
  Hubble horizon cannot affect the density perturbation 
  on super-horizon scales. However, Damour and Mukhanov 
  recently considered the model 
  with the non-convex type potential \cite{DM98}. 
  In this model, taking the time average,  
  the potential energy of the inflaton becomes large compared 
  to the kinetic energy and the accelerated expansion can take 
  place in the oscillating stage. Liddle and Mazumdar called it 
  {\it oscillating inflation} and confirmed this fact numerically
  \cite{LM98}. 
  As was suggested by Damour and Mukhanov, it could be expected
  that the very broad parametric resonance occurs and the quantum 
  fluctuation is enormously amplified. 
  This implies that the adiabatic metric perturbation with the large 
  amplitude might be produced on super-horizon scales in the course of 
  the accelerated expansion, which can lead to the different conclusion 
  from the previous results \cite{KH96,NT97}. Hence, there exists a 
  possibility that the cosmological objects such as the primordial 
  black holes may be formed by the large amplitude of the metric 
  perturbation. 

  In this paper, to explore this possibility,  
  we study the cosmological perturbation in the oscillating inflation 
  and investigate the efficiency of the parametric resonance.  
  We describe the model 
  in Sec.\ref{sec: model}. The parametric resonance 
  of density perturbation is discussed 
  in Sec.\ref{sec: perturb}. 
  We show that the efficiency of the parametric amplification 
  on the spectrum of the curvature perturbation depends on the 
  energy scale of the oscillating inflation and the growth factor.  
  Section \ref{sec: analysis} is devoted to the analysis of 
  the growth factor. Neglecting the metric perturbation, 
  we analytically find that the instability band of the 
  parametric resonance is very broad and the growth factor may 
  become significantly large. We then proceed the 
  numerical calculation of cosmological perturbation without  
  any approximation. The numerical analysis show that the 
  perturbation can experience the parametric amplification, 
  however, the amplification is 
  strongly suppressed when the wavelength of the fluctuation 
  approaches the Hubble horizon scales. The results are 
  briefly summarized and the conclusion in the single field case 
  is described in the last section \ref{sec: conclusion}. 
  We also discuss the oscillating inflation in the multi-field system. 
  It could be expected that the significant amplification of the density 
  perturbation can appear on large scales and may lead to the 
  production of the cosmological black holes.
%
%
%
\section{Oscillating inflation model}
\label{sec: model}
%
%
%
Let us consider the minimally coupled scalar field $\phi$ 
in the flat FRW universe. The homogeneous background equations become
\begin{eqnarray}
  \label{FRW-eq1}
&& 3H^2=\frac{1}{\tilde{M}^{2}_{pl}}
\left(\frac{1}{2}\dot{\phi}^2+V(\phi)\right)~~;
~~~~~\tilde{M}_{pl}=\sqrt{\frac{3}{8\pi}}~M_{pl},
\\
  \label{FRW-eq2}
&& \ddot{\phi}+3H\dot{\phi}+V_{,\phi}=0,
\end{eqnarray}
where $H$ is the Hubble parameter and $M_{pl}$ is the Planck mass.  
%
%
The evolution of the scalar field 
depends on the shape of the potential $V(\phi)$.
We shall consider the model of {\it oscillating inflation} 
proposed by Damour and Mukhanov\cite{DM98}, in which 
the potential is given by
\begin{equation}
  \label{osc-potential}
  V(\phi)=\frac{A}{q}
  \left[\left(\frac{\phi^2}{\phi_c^2}+1\right)^{q/2}-1\right].
\end{equation}
Note that the potential (\ref{osc-potential}) becomes logarithmic, 
$V=(A/2)\log{(\phi^2/\phi_c^2+1)}$ in the limit $q\to0$. 

When the effective mass of the inflaton is larger than the 
Hubble parameter, i.e, $|V_{,\phi\phi}|\gg H^2$, the inflaton field 
shows oscillatory behavior. For the large amplitude $\phi\gg\phi_c$, 
taking the time average during a period of oscillation yields 
the following virial theorem: 
  \begin{equation}
    \label{virial}
    \langle\dot{\phi}^2\rangle \simeq q~\langle V(\phi)\rangle.   
  \end{equation}
Using this relation,  
the inflaton field can be regarded as the perfect fluid matter with 
the equation of state $P=(\gamma_{osc}-1)\rho$, where $\gamma_{osc}$ 
is the effective adiabatic index given by 
\begin{equation}
  \label{osc-index}
  \gamma_{osc}=\frac{2q}{q+2}.
\end{equation}
Thus the time dependences of the scale factor, 
the amplitude of the inflaton field 
and inflaton potential are obtained\cite{DM98}\cite{T83}: 
\begin{eqnarray}
  && a(t)~\propto~t^{(q+2)/3q},
\nonumber 
\\
  && \bar{\phi}(t)~\propto~a^{-6/(q+2)}~\propto~t^{-2/q},
\label{time-dependence}
\\
  && V(\bar{\phi})~\propto~a^{-6q/(q+2)}~\propto~t^{-2},
\nonumber
\end{eqnarray}
where $\bar{\phi}$ is the amplitude of the scalar field evaluated 
at the time that $\dot{\phi}=0$ holds.

Eq.(\ref{time-dependence}) implies that 
  if the potential index $q$ is smaller than unity, the 
 universe can experience the accelerated expansion even in the 
 oscillating phase. 
When the scalar field approaches to the critical value 
$\bar{\phi}\simeq\phi_c$, the 
oscillating inflation ends. 
Hereafter, we shall restrict our attention on 
the inflationary phase with the potential index $q<1$.
%
%
%
%
%
%
\section{Cosmological perturbation}
\label{sec: perturb}
%
%
%
%
%
We now investigate the cosmological perturbation 
  in the oscillating inflation. We shall ignore the 
  interaction with the other scalar fields which leads to the 
  inflaton decay in the reheating process\cite{KLS94,KLS97,GKLS97}.   
  In Ref.\cite{KH96,NT97}, the useful and well-behaved perturbed 
  quantity during the oscillating stage is found :
  \begin{equation}
    Q=\delta{\phi}-\frac{\dot{\phi}}{H}{\cal R}, 
  \end{equation}
  where $\delta\phi$ and ${\cal R}$ are the perturbation of 
  the scalar field and the spatial curvature, respectively \cite{KS84}. 
  The quantity $Q$ is the gauge-invariant variable introduced by 
  Mukhanov\cite{Mukhanov88,MFB92} and 
  the evolution equation is simply given by
  \begin{equation}
    \label{Mukhanov-eq}
    \ddot{Q}+3H\dot{Q}+
    \left[\left(\frac{k}{a}\right)^2+V_{,\phi\phi}+
      6\tilde{M}^{-2}_{pl}\left(\frac{V(\phi)}{H}\right)^.\right]Q=0. 
  \end{equation}
If we neglect the third term in the bracket,   
  the above equation is just reduced to the equation for 
  perturbation $\delta\phi$ ignoring the metric perturbation. 
  We thus understand that the effect of gravitational interaction to 
  the scalar field perturbation is 
  encoded in the third term in the bracket.  

In equation (\ref{Mukhanov-eq}), there exists the exact 
solution in the long-wavelength limit \cite{NT98}. For 
$k\to0$, we have 
  \begin{equation}
    \label{0-mode}
    Q=c_1 \frac{\dot{\phi}}{H}+ c_2 \frac{\dot{\phi}}{H}
    \int \frac{dt}{a^3}\frac{H^2}{\dot{\phi}^2}, 
  \end{equation}
  where $c_1$ and $c_2$ are the integration constants. The solution 
  proportional to the coefficient $c_1$ is named as the growing 
  mode and the one proportional to $c_2$ as the decaying mode.
\footnote{
  It seems that the singular behavior appears in the decaying 
  mode around the zero points of the time derivative of the scalar 
  field. However, the zero points of the 
  denominator in the integral can be canceled out by 
  the zero points of $\dot{\phi}$ in the numerator. Therefore 
  the solution is regular 
  and the amplitude of the solution decreases in time. For more 
  rigorous proof of the regularity and the explicit calculation 
  of the long-wavelength perturbation, see \cite{KH96} and 
  \cite{NT97}.}
%
For the growing mode, the curvature perturbation on comoving 
  slice ${\cal R}_c$ defined by 
  \begin{equation}
    \label{Bardeen-para}
    {\cal R}_c=\frac{H}{\dot{\phi}}Q  
  \end{equation}
  remains constant in time. The curvature perturbation is also referred 
  to as the Bardeen parameter, which is simply related to 
  the CMB temperature fluctuation $\Delta T/T $ 
  observed by the Cosmic Background Explorer satellite \cite{LR98}. 
  In this paper, matching the amplitude of quantum fluctuation $Q$ 
  produced during the oscillating inflation inside the horizon with 
  that of the long wavelength growing mode,  
  we will evaluate the power spectrum 
  of the curvature perturbation ${\cal P}_{{\cal R}}(k)$.  

We focus on the evolution of the quantity $Q$ on sub-horizon scales. 
In equation (\ref{Mukhanov-eq}), the time dependence of 
the terms in the bracket is estimated by using the results 
(\ref{time-dependence}):
  \begin{eqnarray}
   & V_{,\phi\phi}(\bar{\phi})~&\propto~
       a^{-6\left(\frac{q-2}{q+2}\right)},
\label{mass-time}
\\
   & \left.\left(\frac{V(\phi)}{H}\right)^.\right|_{\phi=\bar{\phi}}
      \left.\simeq \frac{V_{,\phi}\dot{\phi}}{H}\right|_{\phi=\bar{\phi}}
      ~&\propto~a^{-6\left(\frac{q-1}{q+2}\right)},
\label{metric-time}
  \end{eqnarray}
which are valid for $\bar{\phi}\gg\phi_c$. During the 
oscillating inflation, the second term in the bracket becomes 
dominant compared to the term (\ref{metric-time}). 
The dominant contribution of the oscillating term (\ref{mass-time}) 
differs from the situation considered in Ref.\cite{KH96}\cite{NT97}.  
We then introduce the following variables:
  \begin{eqnarray}
    &&  Q=\bar{\phi}(t)\tilde{Q}(\eta),
    \nonumber   \\
    &&  d\eta=g(t)dt~~~;~
    g(t)=\frac{\sqrt{A}}{\bar{\phi}}\left(\frac{\bar{\phi}}{\phi_c}\right)^{q/2}
    ~\propto~a^{-3\left(\frac{q-2}{q+2}\right)}.
    \nonumber
  \end{eqnarray}
%
Ignoring the term (\ref{metric-time}), the evolution equation 
(\ref{Mukhanov-eq}) is rewritten as the normal form :
  \begin{equation}
    \label{reduced-mukhanov}
    \tilde{Q}_{,\eta\eta}+
    \left[\left(\frac{k}{ga}\right)^2+
      \tilde{V}_{,\tilde{\phi}\tilde{\phi}}(\tilde{\phi})
      \right]\tilde{Q}=0,
  \end{equation}
where $\tilde{\phi}$ is defined by $\phi(t)=\bar{\phi}~\tilde{\phi}(\eta)$ 
and satisfies the equation of motion 
\begin{equation}
  \label{eom-tilde-phi}
  \tilde{\phi}_{,\eta\eta}+\tilde{V}_{,\tilde{\phi}}(\tilde\phi)=0, 
\end{equation}
with the initial condition $\tilde{\phi}(\eta_i)=1$. 
The quantities $\tilde{V}_{,\tilde{\phi}\tilde{\phi}}$ and 
$\tilde{V}_{,\tilde{\phi}}$ are obtained by differentiating 
the effective potential $\tilde{V}$ with respect to $\tilde{\phi}$: 
  \begin{equation}
    \tilde{V}(\tilde{\phi})=\frac{1}{q}
    \left[\left(\tilde{\phi}^2+\frac{\phi_c^2}{\bar{\phi}^2}\right)^{q/2}-
      \left(\frac{\phi_c}{\bar{\phi}}\right)^q\right].    
      \label{effective-V}
  \end{equation}

Since we have $\tilde{V}\simeq \tilde{\phi}^q/q$,  
$\tilde{\phi}$ oscillates rapidly during the oscillating inflation. 
Treating the time dependence of the term $\tilde{V}$ adiabatically, 
the period of oscillation $\tilde{\phi}$ is 
estimated as 
  \begin{equation}
    T=4\int^1_0\frac{d\tilde{\phi}}{\sqrt{2(q^{-1}-\tilde{V})}}
    \simeq\sqrt{\frac{8\pi}{q}}\frac{\Gamma{(\frac{1}{q})}}
{\Gamma{(\frac{1}{q}+\frac{1}{2})}}. 
\label{period}
  \end{equation}
For the range $0.01\leq q\leq1$, we have $T\simeq5$.

According to the Floquet theorem, presence of the oscillating term 
in (\ref{reduced-mukhanov}) implies that there exists  
the following solution for the equation (\ref{reduced-mukhanov}). 
Since $\tilde{V}_{,\tilde{\phi}\tilde{\phi}}$ is given by $\eta=T/2$, 
we have \cite{AS65}
\begin{equation}
  \label{floquet-sol}
  \tilde{Q}(\tau)=e^{\mu\tau}~P(\tau)~~;~P(\tau+\pi)=P(\tau),
\end{equation}
where we define the new time parameter $\tau\equiv(2\pi/T)\eta$. The 
periodic function $P(\tau)$ is bounded and has a finite amplitude. 
The characteristic exponent $\mu$ takes either a imaginary or a real number, 
which depends on the shape of $\tilde{V}_{,\tilde{\phi}\tilde{\phi}}$ 
and the parameter $k/ga$. The parametric 
resonance means that the characteristic exponent has a real number.

We can see that the parametric resonance 
can affect the curvature perturbation ${\cal R}_c$ produced during 
the oscillating inflation as follows. 
As usual, in the slow-rolling inflation, 
the amplitude of quantum fluctuation is given by 
$Q\simeq|\delta{\phi}|\sim H/2\pi$, which can be deduced from 
the instability of the massless scalar field in de Sitter space
\cite{Linde}. On the other hand, 
the amplification of the quantum fluctuations 
can appear due to the parametric resonance in the oscillating inflation. 
We have 
  \begin{equation}
    \left|\frac{\tilde{Q}_*}{\tilde{Q}_0}\right|\simeq e^{\mu\Delta\tau},
    \label{amplify-Q}
  \end{equation}
  where subscripts $(_0)$ and $(_*)$ denotes the quantities evaluated 
  at the beginning of oscillating inflation and at the time when the 
  wavelength of the quantum fluctuation exceeds the Hubble horizon scale 
  $(k=a_*H_*)$, respectively. Here, $\Delta\tau$ denotes the time interval 
  during which the quantum fluctuation is amplified by the parametric 
  resonance.
For the modes inside the horizon $k/aH\to\infty$, the mode function 
  of the equation (\ref{Mukhanov-eq}) initially becomes 
  $a_0Q_0\to \exp{(i\int dt/a)}/\sqrt{2k}$, corresponding to the 
  vacuum state in the Minkowski spacetime\cite{SL93}. 
  Matching the amplitude of the short wavelength fluctuation 
  with that of the long wavelength solution (\ref{0-mode}), 
  the curvature perturbation (\ref{Bardeen-para}) 
  is roughly estimated : 
  \begin{equation}
    |{\cal R}_c|\simeq \left(\frac{H}{\dot{\phi}}\right)_*
    \frac{H_*}{\sqrt{2k^3}}\left(\frac{\bar{\phi}_*}{\bar{\phi}_0}\right)
    ^{(4-q)/(2+q)}\left|\frac{\tilde{Q}_*}{\tilde{Q}_0}\right|,
  \end{equation}
which is evaluated at the horizon-crossing time.  
Therefore, the power spectrum of the curvature perturbation 
${\cal P}_{\cal R}(k)$ evaluated by taking the ensemble average 
$\langle\cdots\rangle_{ens}$ becomes
\begin{equation}
  \label{power-spectrum}
  {\cal P}^{1/2}_{{\cal R}}(k)\equiv\sqrt{\frac{k^3}{2\pi^2}}
  \langle|{\cal R}_c|^2\rangle^{1/2}_{ens}
  \simeq \frac{\gamma_{osc}^{-1}}{2\pi}
  \left(\frac{\bar{\phi}_*}{\bar{\phi}_0}\right)^{(4-q)/(2+q)}
  \left(\frac{V_*^{1/4}}{\tilde{M}_{pl}}\right)^2~ e^{\mu\Delta{\tau}}. 
\end{equation}
In the derivation of the expression (\ref{power-spectrum}), 
we have naively replaced the ensemble average 
$\langle\dot{\phi}^2\rangle_{ens}$ with the time average 
$\langle\dot{\phi}^2\rangle$ and used the virial theorem (\ref{virial}). 
Eq.(\ref{power-spectrum}) shows that the appearance of exponential 
factor $e^{\mu\Delta\tau}$ may give the important contribution to the 
spectrum of curvature perturbation, although the amplitude of 
${\cal P}_{\cal R}$ is always suppressed by the energy scale of 
the oscillating inflation $V_*^{1/4}\equiv V^{1/4}(\bar{\phi}_*)$  
\cite{SL93}\cite{LR98}. 
%
%
%
\section{Parametric Amplification during oscillating inflation}
\label{sec: analysis}
%
%
%
In the previous section, we have seen that the effect of 
parametric resonance can lead to the amplification of 
the curvature perturbation, which may gives ${\cal P}_{\cal R}\sim1$. 
In this section, to explore the efficiency of the parametric 
amplification, we analytically and numerically investigate 
the growth factor $e^{\mu\Delta\tau}$.
%
%
%
\subsection{Analytic estimation}
%
%
%
  Let us evaluate the characteristic exponent $\mu$ from 
  the reduced equation (\ref{reduced-mukhanov}). 
  Using the solution (\ref{floquet-sol}) obtained from the 
  Floquet theorem, we have the following formula \cite{AS65}:
\begin{equation}
  \mu=\frac{1}{\pi}\log{\left[\sqrt{F^2}+\sqrt{F^2-1}\right]}~~~;~~
  F=1+2\tilde{Q}_1'(\tau=\frac{\pi}{2})
  \tilde{Q}_2(\tau=\frac{\pi}{2}),
  \label{floquet-formula}
\end{equation}
where the prime denotes the differentiation with respect to the time $\tau$. 
$\tilde{Q}_1$ and $\tilde{Q}_2$ are the solutions satisfying 
the initial conditions $\tilde{Q}_1=1$, $\tilde{Q}'_{1}=0$ and 
$\tilde{Q}_2=0$, $\tilde{Q}'_{2}=1$ at $\tau=0$, respectively.
The parametric resonance appears in the case $F>1$. 

Although it is difficult to obtain the solutions $\tilde{Q}_1$ and 
$\tilde{Q}_2$ for our complicated potential 
$\tilde{V}_{,\tilde{\phi}\tilde{\phi}}$, we can get the approximate 
expression of the characteristic exponent. 
The effective mass of the oscillating inflaton field becomes 
  negative at the shallow wings of the potential and positive at the 
  minimum. Thus we replace the periodic function 
  $\tilde{V}_{,\tilde{\phi}\tilde{\phi}}$ with the {\it step-wise} function 
  given by 
\begin{equation}
\tilde{V}_{,\tilde{\phi}\tilde{\phi}}\longrightarrow
\left\{
\begin{array}{c}
\tilde{V}_{,\tilde{\phi}\tilde{\phi}}(\tilde{\phi}=0), ~~~~~~ (|\tau|<\tau_1) 
\\
\\
\tilde{V}_{,\tilde{\phi}\tilde{\phi}}(\tilde{\phi}=1),  ~~~~~~~
(\tau_1<|\tau|<\frac{\pi}{2}) \\
\end{array}
\right.,
  \label{replace-of-V}
\end{equation}
 where we choose 
\begin{eqnarray}
&&  \tau_1=\frac{2\pi}{T}\int_0^{\tilde{\phi_1}}
\frac{d\tilde{\phi}}{\sqrt{2(q^{-1}-\tilde{V})}}~~~;~~
\tilde{\phi}_1=\sqrt{\frac{3}{1-q}}\frac{\phi_c}{\bar{\phi}}.
\nonumber
\end{eqnarray}
At $\tilde{\phi}=\tilde{\phi}_1$, 
we have $\tilde{V}_{,\tilde{\phi}\tilde{\phi}\tilde{\phi}}=0$. 
The approximation (\ref{replace-of-V}) 
enables us to obtain the analytic solutions $\tilde{Q}_1$ and 
$\tilde{Q}_2$. 
Then the quantity $F$ is evaluated as follows:
\begin{equation}
  \label{F}
  F=\frac{1}{2}\left(\frac{\alpha}{\beta}+\frac{\beta}{\alpha}\right)
  \sin{(2\alpha\tau_1)}\sin{(2\beta(\tau_1-\frac{\pi}{2}))}
  +\cos{(2\alpha\tau_1)}\cos{(2\beta(\tau_1-\frac{\pi}{2}))},
\end{equation}
where 
\begin{eqnarray}
  &&
\alpha=\frac{T}{2\pi}\left[
\left(\frac{k}{ga}\right)^2+\tilde{V}_{,\tilde{\phi}\tilde{\phi}}
(\tilde{\phi}=0)\right]^{1/2},~~~~~
\beta=\frac{T}{2\pi}\left[
\left(\frac{k}{ga}\right)^2+\tilde{V}_{,\tilde{\phi}\tilde{\phi}}
(\tilde{\phi}=1)\right]^{1/2}.
\nonumber
\end{eqnarray}
Since $\tilde{V}_{,\tilde{\phi}\tilde{\phi}}(\tilde{\phi}=1)<0$ 
during the oscillating inflation, the variable $\beta$ becomes imaginary 
for the long-wave modes $(k/a)^2\lesssim|V_{,\phi\phi}(\bar{\phi})|$ 
and we obtain $F>1$. This can be deduced from the negative coupling 
instability\cite{DM98}. 
Important point is that we could also have $F>1$ for the short-wave mode 
$(k/a)^2\gtrsim|V_{,\phi\phi}(\bar{\phi})|$. 
In the latter case, the band structure of the characteristic exponent 
$\mu$ appears. 

In Fig.1, we plot the characteristic exponent $\mu$ 
as a function of $k/ga$ 
evaluated from (\ref{floquet-formula}) and (\ref{F}).  
The parameters are chosen as 
$\phi_c=10^{-6}~\tilde{M}_{pl}$ and $q=0.1$. As for 
the quantity $\bar{\phi}$, we treat it as constant and set 
$\bar{\phi}=q~\tilde{M}_{pl}/\sqrt{6}$, 
corresponding to the initial value of the oscillating inflation
\cite{LM98}. The circles 
show the characteristic exponent obtained from the 
substitution of the numerical solutions of equations 
(\ref{reduced-mukhanov}) and (\ref{eom-tilde-phi}) into the 
formula (\ref{floquet-formula}). We see that 
the approximation predicts the numerical result reasonably well. 
The crucial observation is that the instability band is very broad 
and the characteristic exponent can be of the order of unity. 

To evaluate the growth factor $e^{\mu \Delta}$, we further need 
the period $\Delta\tau$. It is given by 
\begin{equation}
  \label{D-tau}
  \Delta\tau=\frac{2\pi}{T}\int dt g(t)=\sqrt{\frac{\pi}{72}}~q(q+2)
\frac{\Gamma(\frac{1}{q}+\frac{1}{2})}{\Gamma(\frac{1}{q})}
\left(\frac{\tilde{M}_{pl}}{\bar{\phi}_*}-
\frac{\tilde{M}_{pl}}{\bar{\phi}_0}\right),
\end{equation}
where the integral is evaluated from 
the time at which the oscillating 
inflation starts to the horizon-crossing time. Eq.(\ref{D-tau}) 
is valid for the case $q>0$.  
Rewriting the above equation with the relation $k=a_*H_*$, 
the wave number dependence of the growth factor $e^{\mu\Delta\tau}$  
shows that the characteristic peak appears on the spectrum of 
curvature perturbation (\ref{power-spectrum}). 
Because of the rapid oscillation of the inflaton field, the period 
$\Delta\tau$ effectively becomes very large. For $q=0.1$, 
the typical values $\bar{\phi}_*=10^{-4}$, $10^{-5}$ 
and $10^{-6}~\tilde{M}_{pl}$ give $\Delta{\tau}=1367$, 
$13696$ and $136991$, respectively. 
Therefore, even if we get the rather small value of the 
characteristic exponent $\mu=0.01$, for example, 
the contribution of the growth factor to the curvature perturbation 
becomes significantly large.  For $\bar{\phi}_*=10^{-5} \tilde{M}_{pl}$ 
and $q=0.1$, we have $e^{\mu\Delta\tau}=10^{182}$ !  
%
%
%
\subsection{Numerical result}
%
%
%
The analytic estimation in the previous subsection 
  rather overestimates the growth factor. 
  For more rigorous evaluation, we must 
  take into account the time dependence of $\bar{\phi}$ and 
  $k/ga$. 
Furthermore, we should remember that the previous result 
  comes from the analysis of the reduced 
  equation (\ref{reduced-mukhanov}). 
  This also leads to the overestimation of the growth factor. 
  However, there still exists a possibility of the parametric 
  amplification of the curvature perturbation spectrum. 
  To clarify the influence of parametric resonance, 
  we numerically solve the equations (\ref{FRW-eq1}), (\ref{FRW-eq2}) 
  and (\ref{Mukhanov-eq}). 

Fig.2 shows the growth factor $e^{\mu\Delta\tau_e}$ for the index $q=0.1$, 
i.e, the ratio of the amplitude $\tilde{Q}_e$ evaluated 
at the end of oscillating 
inflation to the amplitude $\tilde{Q}_0$  
at the beginning of inflation by varying the parameter $\phi_c$. 
For each value $\phi_c$, 
we start to calculate the background evolution by setting  
the value $\phi_0=q \tilde{M}_{pl}/\sqrt{6}$ corresponding to 
the final value of the slow-rolling inflation. 
As for the fluctuations, the initial conditions $\tilde{Q}'_0=0$ are chosen. 
Although we have checked the evolution for the various initial conditions 
$\tilde{Q}_0\propto \cos{(k\int dt/a)}$, corresponding to the 
vacuum state in the short wavelength limit, the maximal amplitude can be 
obtained from the initial condition $\tilde{Q}'_0=0$.  
Fig.2 shows that the amplification for the fluctuations occurs 
by the effect of parametric resonance.  
It is obvious that the maximal amplitude depends on 
the parameter $\phi_c$. For the smaller value of $\phi_c$, the 
amplitude of the fluctuation becomes larger, which comes 
from the fact that the period of inflation represented by 
the e-folding number $N_e=\log_e{(a_e/a_0)}$ depends on 
the critical value $\phi_c$ \cite{DM98,LM98}. When the oscillating 
inflation takes place for a longer time, the effect of parametric 
resonance works out more efficiently and the fluctuations are significantly 
amplified. In Fig.2, the three different lines are depicted. 
The solid line is the fluctuation with the wavelength ten times 
smaller than the radius of the Hubble horizon at the end 
of oscillating inflation, i.e, $k=10a_eH_e$. 
The dashed line has the wavelength 
$k=2a_eH_e$ and the dotted line corresponds to the fluctuation 
whose wavelength just reaches at the Hubble radius after oscillating 
inflation. 
It is remarkable that the modes inside the Hubble horizon 
are enormously amplified, while the amplification of the long-wave mode 
comparable to the Hubble horizon is strongly suppressed. Though 
the dotted line apparently takes the growth factor $\sim10^7$, 
it can be recognized that the parametric resonance becomes 
ineffective for the long-wave modes. 

To see the efficiency of the parametric amplification explicitly, 
we plot in Fig.3 the time evolution of the fluctuations using 
the quantity $Q$ from the beginning to the end of oscillating 
inflation. We set the parameters 
$q=0.1$ and $\phi_c=10^{-5}\tilde{M}_{pl}$. 
The horizontal axis represents 
the cosmic time normalized by the factor $\sqrt{A}/\tilde{M}_{pl}$.
Each figure shows the time evolution 
with the different wavelength under the same initial conditions 
$Q_0=1.0$ and $Q_0'=0$ : (a) $k=10a_eH_e$;(b) $k=2a_eH_e$;
(c) $k=a_eH_e$. These figures  clearly reveal that the 
growth of the perturbation is much 
sensitive to the wavelength of the fluctuations. 
The parametric amplification works efficiently inside the 
horizon, however, the amplitude $Q$ becomes constant for the 
wavelength near the Hubble horizon. 

We can understand these behavior as follows. As was 
described in Sec.\ref{sec: perturb}, there exists the 
exact solution for the variable $Q$ in the limit $k\to0$. 
Ignoring the decaying mode proportional to the coefficient $c_2$ in 
(\ref{0-mode}), the amplitude of the long wavelength solution  
becomes nearly constant, which can be deduced 
from (\ref{time-dependence}). This behavior can also be obtained in the 
case of $k\neq0$, when the term $(k/a)^2$ in equation (\ref{Mukhanov-eq})
 is negligible compared to the 
terms $V_{\phi\phi}$ and $\tilde{M}^{-2}(V/H)^\cdot$, which gives 
the condition $k/a\ll H$ \cite{NT97}. In the previous subsection, we 
analyzed the growth factor $e^{\mu\Delta\tau}$ for the short-wave mode 
by dropping the term $\tilde{M}^{-2}(V/H)^\cdot$. 
Fig.3 indicates that the term induced by the gravitational perturbation 
plays a crucial role even for the modes comparable to the Hubble horizon. 
This is fully consistent with the result of the paper 
\cite{KH96} and \cite{NT97}.  
Therefore, the spectrum of the curvature perturbation 
produced during oscillating inflation on super-horizon scales 
has the rather small amplitude compared with the analytic prediction. 
%
%
%
\section{Concluding remarks}
\label{sec: conclusion}
%
%
%
We have analyzed the cosmological perturbation in the 
oscillating inflation 
and explored a possibility of parametric amplification for   
the curvature perturbation. 
Neglecting the metric perturbation, 
the analytic estimation shows that the curvature perturbation could be 
amplified by the broad band parametric resonance. 
We then numerically examined this issue by solving the 
evolution equation for perturbation without any approximation. 
The enormous amplification of the perturbation 
can appear, however, we found that the presence of the 
the metric perturbation strongly suppresses  
the parametric amplification on large scales comparable to the 
Hubble horizon.  

Now we discuss the cosmological implication of the 
parametric resonance to the density perturbation. 
According to Ref.\cite{LL93,LR98}, 
the physical length of the fluctuation produced during the  
oscillating inflation is 
\begin{equation}
  \lambda_{phy}~\lesssim ~10^{1+0.43 \alpha_*}~~\mbox{cm},
  \label{cosmological-scale}
\end{equation}
where $\lambda_{phy}$ denotes the present wavelength. The quantity 
$\alpha_*$ is given by 
\begin{equation}
  \alpha_*=~N_*+\log_e{\left(\frac{10^{16}~\mbox{GeV}}{V_*^{1/4}}\right)}.
  \label{alpha}
\end{equation}
$N_*$ denotes the e-folding number of the oscillating inflation 
evaluated at the Hubble crossing time. 
On the other hand, if the amplitude ${\cal R}_c$ produced during 
the inflation 
becomes of the order of unity, the fluctuations can experience the 
  gravitational collapse. Assuming that the primordial black 
  holes are formed at the horizon re-entry time during the 
  radiation epoch,  we can evaluate the typical mass of a 
  black hole using the equation (\ref{cosmological-scale}). 
  We obtain
\begin{equation}
  M_{BH}~\lesssim~10^{1+0.86\alpha_*}~~\mbox{g}, 
  \label{pbh-mass}
\end{equation}
which is evaluated at the formation epoch. 

As is shown by Liddle and Mazumdar \cite{LM98}, 
a period of oscillating inflation is short. This can be 
  checked in our numerical calculation.  For the critical value 
  $\phi_{c}=10^{-6}\tilde{M}_{pl}$, we obtain the small value of the 
  e-folding number $N_{e}=3.5$. 
Hence, the expressions  
  (\ref{cosmological-scale}) and (\ref{pbh-mass}) show that 
  the energy scale of the oscillating inflation should be 
  rather low to have the cosmologically interesting scale for  
  the back hole mass and the physical size of the fluctuations. 
Asaka, Kawasaki and Yamaguchi \cite{AKY98} investigated the 
inflation model to resolve the cosmological moduli problem, 
in which the oscillating 
  inflation occurs at the energy scale $V_{e}^{1/4}\simeq 10 \mbox{GeV}$, 
  instead of the thermal inflation. In this energy scale, 
  if we have the e-folding number $N_e \simeq 3$, it is possible to 
  get $M_{BH}\sim 1 M_{\odot}$, as a candidate for explaining the recent 
  observation of the massive compact halo objects, 
  although the physical length is still small, $\lambda_{phy}\sim 1~pc$.  

However, the amplitude of the vacuum fluctuation 
  becomes extremely small for the inflation at low energy scale. 
  We saw in Sec.\ref{sec: perturb} that the spectrum of curvature 
  perturbation ${\cal P}_{\cal R}$ given by (\ref{power-spectrum}) 
  contains the suppression factor $(V_*^{1/4}/M_{pl})^2$,    
  which takes $\sim 10^{-36}$ in the above case. 
Furthermore, our results show that 
  the gravitational interaction plays a crucial role for the 
  parametric amplification of the curvature perturbation. 
  To obtain the relevant growth factor 
  $e^{\mu\Delta\tau}$ to compensate the suppression 
  factor, a long period of oscillating inflation is necessary, which 
  requires the very fine-tuning of the model parameter 
  $\phi_c$. 
  Therefore, in the single field case, we conclude that the influence of the 
  parametric resonance on the cosmological perturbation 
  would not be imprinted on the universe observed 
  as long as the adiabatic mode of the perturbation is dominant. 

  The conclusion might be changed if we consider the 
  multi-field system. Consider the hybrid-type inflation 
  which is a more realistic model motivated by the particle 
  physics\cite{LR98}. In this scenario, 
  we can imagine that the oscillating inflation occurs 
  subsequent to the slow-rolling inflation and it is 
  followed by a secondary inflation driven by another scalar field.  
  If we choose the appropriate duration of the secondary inflation, 
  the oscillating inflationary phase could have the 
  relevant length scales $\lambda_{phy}$ which can affect the large 
  scale structure. 
  In addition, the hybrid inflation would make a 
  modification of the spectrum ${\cal P}_{\cal R}$. 
  In equation (\ref{power-spectrum}), we should replace the term 
  $(V_*^{1/4}/\tilde{M}_{pl})^2$ with $V_*^{1/4}/\tilde{M}_{pl}$ 
  \cite{RSG96}. 
  This relaxes the fine-tuning of the model parameter $\phi_c$ 
  to give the cosmologically interesting scale of the black hole 
  mass $M_{BH}$.   
  The effect of oscillating inflation might be imprinted on the 
  primordial density perturbation in such inflationary scenario. 
  To investigate the spectrum ${\cal P}_{\cal R}$ of the hybrid inflation,  
  we must study the cosmological 
  perturbations with multi-field system, in which the contribution of 
  iso-curvature perturbation cannot be neglected \cite{TN98}. 
  The non-linear interaction of the oscillating inflaton 
  with the secondary inflation-driven field would lead to the 
  significant amplification of the iso-curvature perturbation 
  by the parametric resonance.  
  Although the influence of iso-curvature mode on the gauge-invariant 
  perturbation ${\cal R}_c$ has not been understood completely, 
  the result in this paper would shed light on the evaluation of 
  the primordial power spectra. We will report the analysis of 
  the multi-field system in a separate publication. 
%
%
%
\acknowledgements
{
The author would like to thank M.Sakagami for providing him 
the numerical code and useful comments, J.Soda for 
careful reading of the manuscript, S.Kawai and K.Koyama 
for valuable discussions and comments. 
}
%
%
%
%
%
%

%
\newpage
%
%
\section*{Figure caption}
\begin{description}
\item[Fig1] Characterisctic exponent $\mu$ for the parameters 
$q=0.1$ and $\phi_c=10^{-6}\tilde{M}_{pl}$. We plot the exponent 
$\mu$ as a function of the Fourier mode $k/ga$. 
The solid line shows the 
approximation using the result (\ref{F}). The circles are 
the characteristic exponent evaluated by substituting 
the numerical solutions of eqs.(\ref{reduced-mukhanov}) and (\ref{eom-tilde-phi}) 
into the formula (\ref{floquet-formula}). In both cases, the variable 
$\bar{\phi}$ in the effective potential $\tilde{V}$ is 
specified as $\bar{\phi}=q\tilde{M}_{pl}/\sqrt{6}$.

\item[Fig2] The growth factor $e^{\mu\Delta\tau_e}$, 
the ratio of the amplitude $\tilde{Q}_e$ evaluated at the end 
of the oscillating inflation to the one $\tilde{Q}_0$ at the beginning of 
inflation. We plot the ratio of the amplitude with the same 
initial condition $\tilde{Q}_0'=0$ by varying the model 
parameter $\phi_c$. We set the potential index $q=0.1$. The solid line 
corresponds to the fluctuation with the mode $k=10 a_eH_e$, i.e, 
the wavelength is ten times smaller than the Hubble horizon size 
at the end of oscillating inflation. The dashed line represents the 
amplitude for the modes $k=2 a_eH_e$. The dotted line is the ratio for  
the fluctuation whose wavelength reaches at the Hubble horizon size 
just after the oscillating inflation ends. 

\item[Fig3] Evolution of $Q$ in the case of the 
parameters $q=0.1$ and $\phi_c=10^{-5} \tilde{M}_{pl}$. For each figure,  
we start to calculate the background eqs.(\ref{FRW-eq1}) and (\ref{FRW-eq2}) 
from the value $\phi=q/\sqrt{6}$ after slow-rolling regime. 
The initial conditions for the 
fluctuation $Q$ are set by $Q=1.0,~Q'=0$ and eq.(\ref{Mukhanov-eq}) 
is solved numerically. The horizontal axis denotes the cosmic time 
normalized by $\sqrt{A}/\tilde{M}_{pl}$: 
(a) The fluctuation $Q$ for the mode $k=10a_eH_e$ ;
(b) The fluctuation $Q$ for the mode $k=2a_eH_e$  ; 
(c) The gauge-invariant quantity $Q$ 
whose wavelength reaches at the Hubble horizon size just after the 
oscillating inflation ends, i.e, $k=a_eH_e$. 
\end{description}
%
%
%
%
%
%
%
%
%
\newpage
%
%
%
\begin{figure}[hbtp]
  \begin{center}
    \epsfysize=9cm
    \leavevmode
    \epsffile{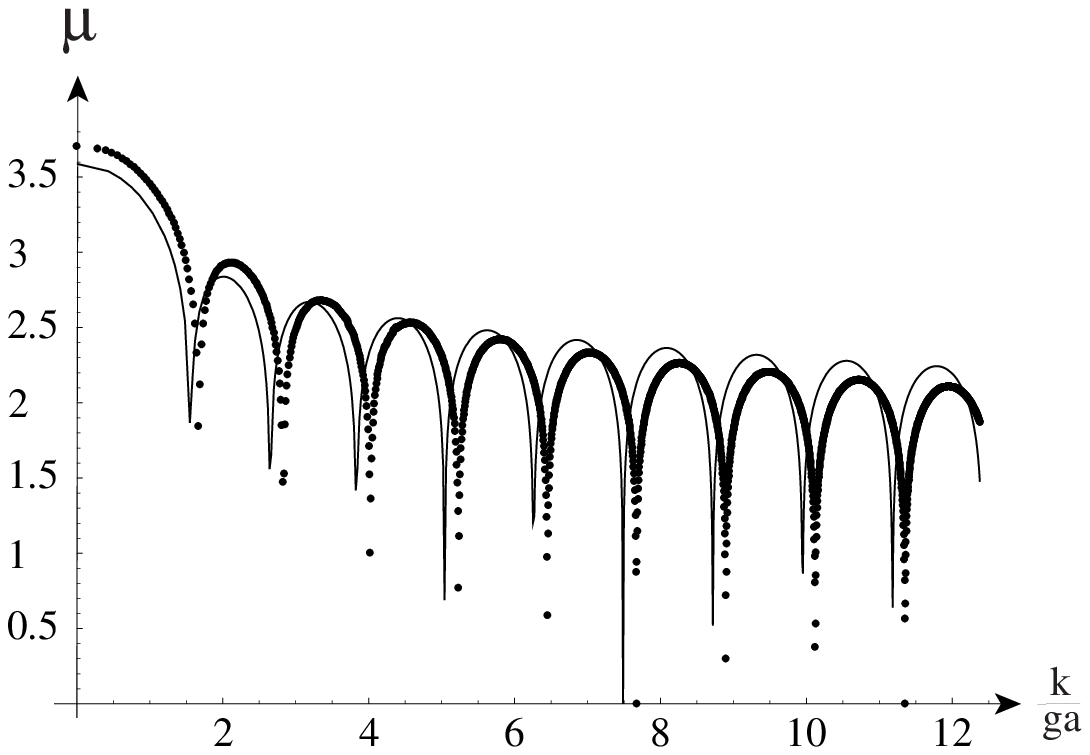}        
    \label{fig:fig1}
  \end{center}
\end{figure}
\vspace{1cm}
\begin{center}
{\large Fig.1}
\end{center}
%
%
%
%
\begin{figure}[hbtp]
  \begin{center}
    \epsfysize=8cm
    \leavevmode
    \epsffile{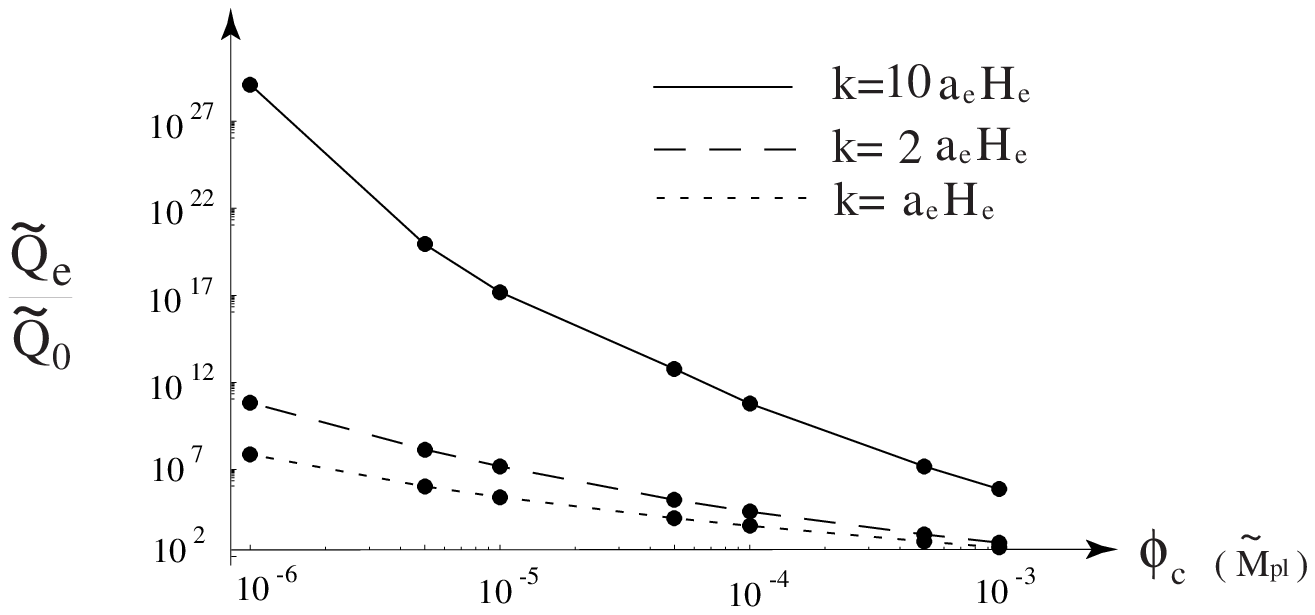}        
  \end{center}
\end{figure}
\begin{center}
{\large Fig.2}
\end{center}
\newpage
\begin{figure}[hbtp]
  \begin{center}
    \epsfysize=6cm
    \leavevmode
    \epsffile{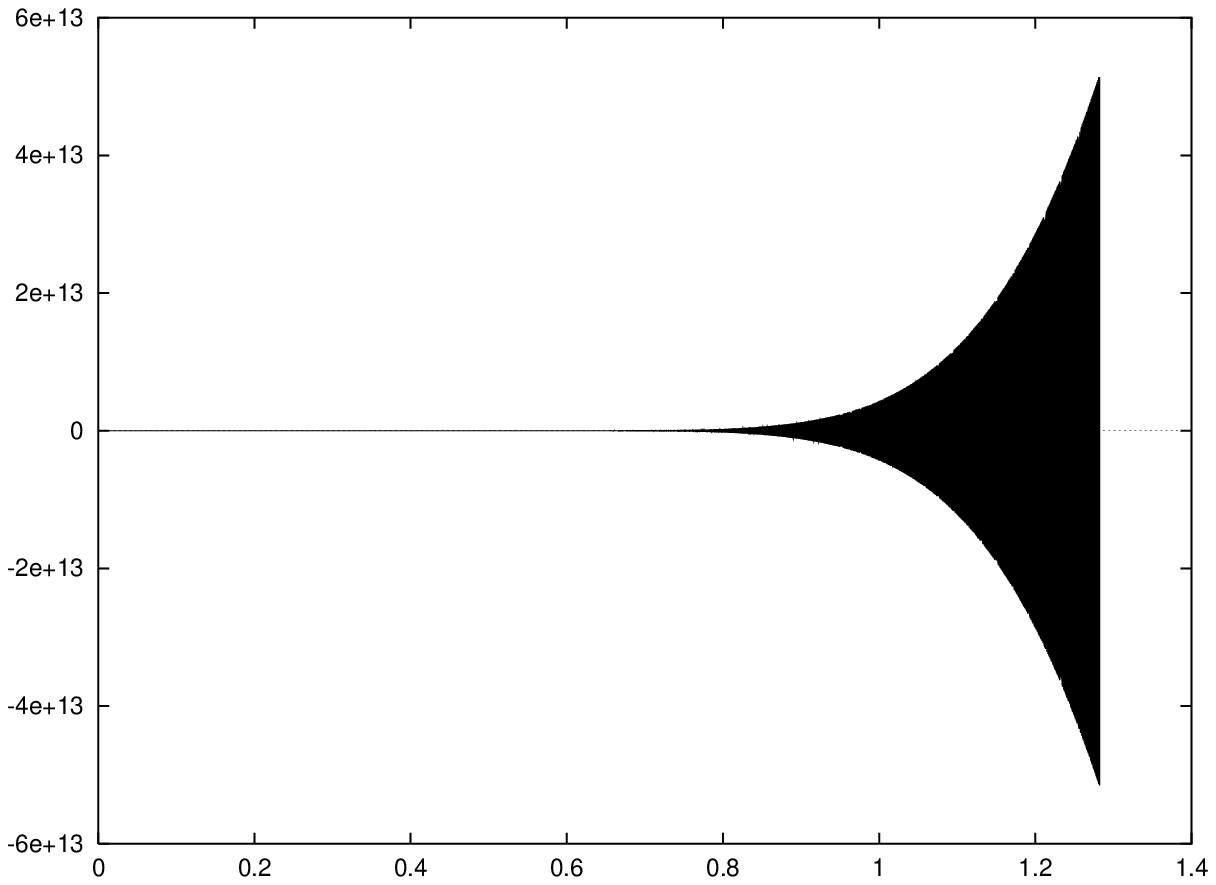}        
  \end{center}
\end{figure}
\begin{center}
{\large Fig.3a}
\end{center}
%
%
\begin{figure}[hbtp]
  \begin{center}
    \epsfysize=6cm
    \leavevmode
    \epsffile{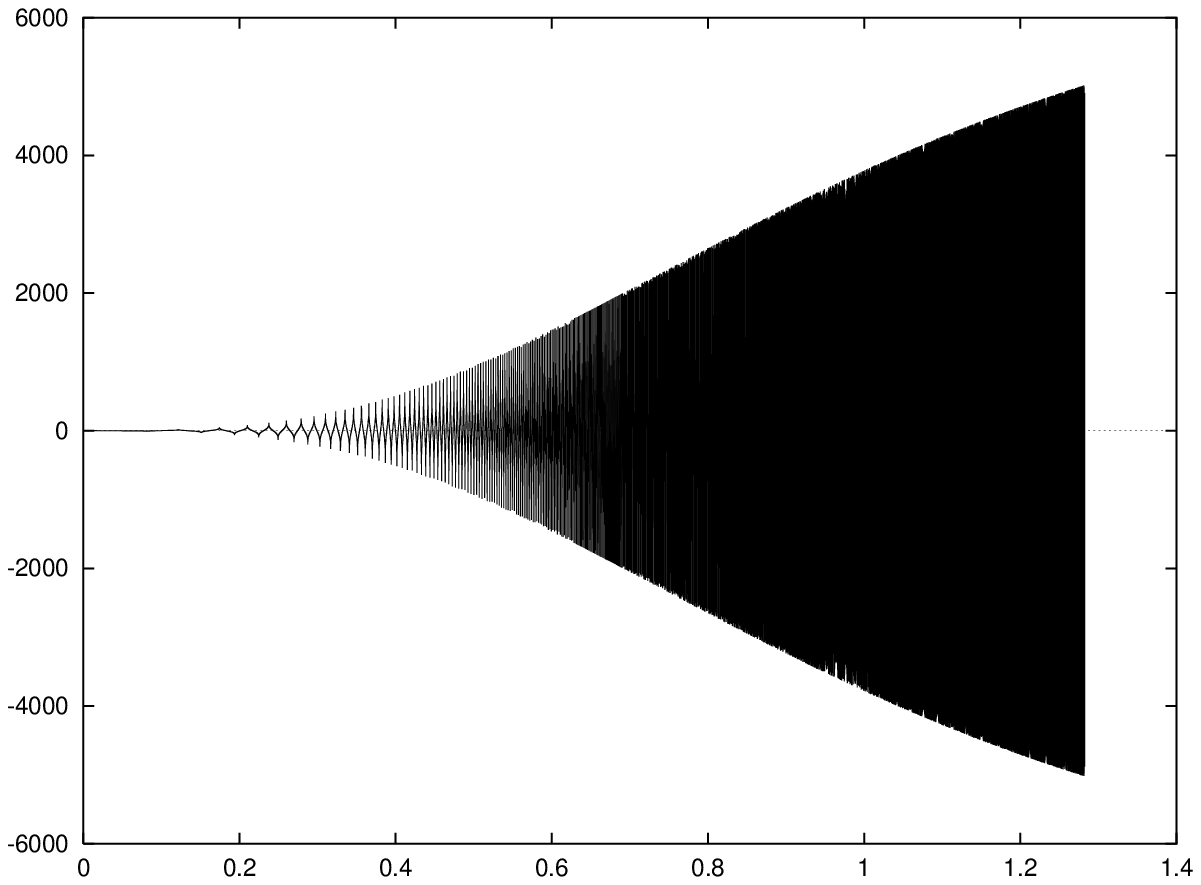}        
  \end{center}
\end{figure}
\begin{center}
{\large Fig.3b}
\end{center}
%
%
\begin{figure}[hbtp]
  \begin{center}
   \epsfysize=6cm
    \leavevmode
    \epsffile{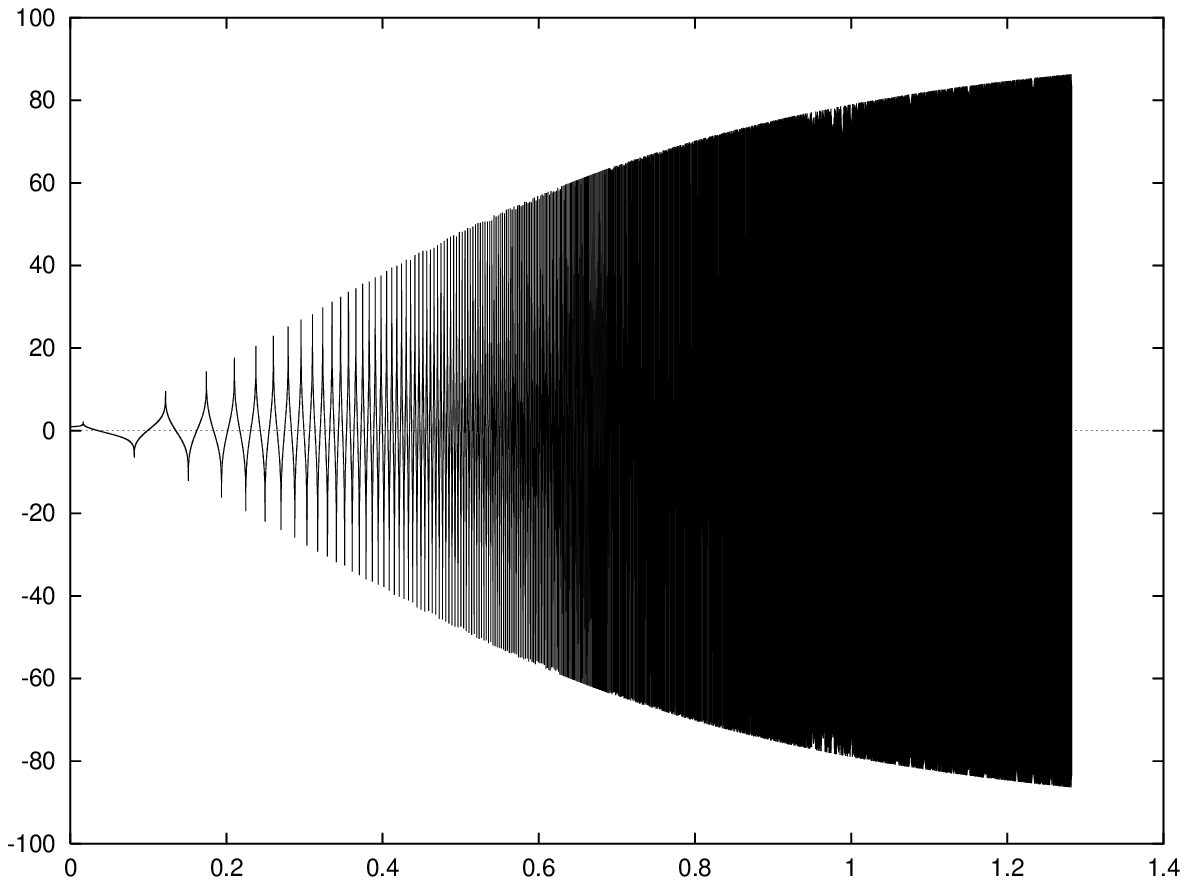}        
  \end{center}
\end{figure}
\begin{center}
{\large Fig.3c}
\end{center}
\end{document}